# Intriguing electronic and optical prospects of FCC bimetallic two-dimensional heterostructures: epsilon near-zero behaviour in UV-vis range

Tuhin Kumar Maji[1], Kumar Vaibhav[2], Ranjit Hawaldar[3], K. V. Adarsh[4], Samir Kumar Pal[1] and Debjani Karmakar[5,*]

[1] Department of Chemical Biological and Macromolecular Sciences, S.N. Bose National Centre for Basics Sciences, Salt Lake, Sector 3, Kolkata 700106, India

[2] Computer Division, Bhabha Atomic Research Centre, Trombay 400085, India

[3] Centre for Materials for Electronics Technology, Off Pashan Road, Panchwati, Pune-411008, India

[4] Department of Physics, Indian Institute of Science Education and Research, Bhopal 462066, India

[5] Technical Physics Division, Bhabha Atomic Research Centre, Trombay 400085, India

*Corresponding author:
Debjani Karmakar: debjan@barc.gov.in

**Abstract**

Higher superconducting critical temperature and large-area epsilon-near-zero systems are two long-standing goals of scientific community, having explicit relationship with chemistry of correlated electrons in localized orbitals. Motivated by the recent experimentally observed strongly correlated phenomena in nanostructures of simple Drude systems, we have theoretically investigated some potential bimetallic FCC combinations starting from large-area interface to embedded and doped two-dimensional (2D) nanostructures, having resemblance with the experimentally studied systems. Using different first-principles techniques, encompassing density functional theory (DFT), time-dependent DFT (TDDFT), phonon and DFT-coupled quantum transport, we propose the following strongly correlated prospects of potential bimetallic nanostructures like Au/Ag and Pt/Pd : 1) For 2D doped and embedded nanostructures of these systems, DFT-calculated non-trivial band-structure and Fermi-surface topology may be emblematic to the presence of instabilities like charge density waves; 2) the optical attributes extracted from the TDDFT calculations for these systems indicate interfacial morphology induced band-localization leading to near-zero behavior of both real and imaginary parts of the dynamical dielectric response in the ultra-violet to visible (UV-vis) optical range; 3) Low-energy intra-band plasmonic oscillations, as present for individual metallic surfaces, are completely suppressed for embedded and doped nanostructures manifesting the sole importance of inter-band transitions, as observed from the TDDFT-derived electron-energy loss spectra; 4) Phonon-dispersion of the nanostructured systems indicates the presence of soft phonons and dynamical instabilities; 5) Quantum transport calculations on simplest possible device made out of these bimetallic systems reveal generation of highly transmitting pockets over the cross-sectional area for some selected device geometry. We envisage that, if scrutinized experimentally, such systems may unveil many fascinating aspects of orbital chemistry, physics and optics promoting relevant applications in many diverse fields.

**Keywords:** Density Functional Theory, Phonons, Optics, Optical materials.

## Introduction

Current amelioration on correlated properties of bimetallic nanostructures (NS) [1-4] has instigated lot of activities in multi-revisited phoenix field, superconductivity, with concurrent genesis of controversies. The first and foremost impact of such studies is, to ignite the expectation to accomplish the long-prophesized goal of near-room temperature superconductivity. The theoretical scrutiny towards the experimental behavior have provided quite contrasting perspective towards the observed phenomenon, *viz.*, electron-transfer induced strong perturbation in monovalent matrix possibly leading to granular superconductivity[5], percolative transition of the disjoint components of the NS attaining a zero-resistive state[6], presence of van-hove singularities due to oxidative reactions[7] or prediction of superconducting $T_C$ calculated for 3D crystals and 2D slabs of Au-Ag alloys in mK range[8].

The experimental system consists of a nanostructure of Ag-nanoparticles of ~ 1nm diameter embedded in Au [111] matrix [1-4]. While the electrical properties of the experimental two-dimensional system has manifested a significant drop in resistivity close to room temperature along with the associated diamagnetism [2,3], the optical property displays a suppression of surface plasmon resonance of the component noble metal systems Au or Ag [1]. This system poses a challenge to the general understanding of physical chemistry, where a nanostructure comprised of two normal Drude metals possessing delocalized itinerant band-electrons, has culminated some experimental results, which should be an outcome of occurrence of localized correlated electronic bands in the system. Therefore, these experimental results provide an urge and motivation to study the static and time-dependent electronic, phonon and transport behavior of Au/Ag nanostructures and large-area interfaces as a result of its structural chemistry.

In former literature, numerous combinations of noble metals are used for plasmonic circuits [9-12] or for designing of metamaterials. Artificially engineered metamaterials are widely used for designing newer material properties, different from the inherited attributes of the consisting components, for innovative applicability in electromagnetic[13], microwave[14] and optical[15] purposes. In addition, being very common systems, bimetallic large-area interfaces and alloy formation have long since been investigated for Au and Ag. Solubility of wide varieties of different shapes of Au-clusters starting from spheres[16], nanorods [17] *etc*. within other metallic matrices and inspection of thermodynamically stable phase diagrams are carried out by adopting both theoretical and experimental routes. Thorough scrutiny of mechanical stability after calculation the interface fracture energy of binary metallic systems are also exercised[18]. For different orientation combinations of interfacial systems, energetics[19] and vibrational properties[20] of epitaxial growth of Au on differently cleaved Ag surfaces are examined[19]. The structural formation of binary clusters and alloys of Au-Ag system in various 2D and 3D morphologies and their respective electronic structure[21] as well as different temperature-composition phases[22] are surveyed. Such bimetallic combinations are also known to have applications in catalytic activity[23] and as surface-enhanced Raman scattering probes[17]. Cercellier *et. al*. had demonstrated experimentally an increase of spin-orbit coupling at the Au/Ag(111) interface with annealing induced disordered alloy formation [24]. Albeit there are copious studies on structural and electronic properties on Au-Ag systems, very few studies concentrated on the correlated electronic behavior at their nanostructured interfaces. The current study attempts to fill the void after addressing the question about the origin of correlated orbital chemistry, its interconnection with nanostructured morphology and the combined impact on the electronic, optical, phonon and transport properties.

In this work, with an extensive electronic structure calculations using density functional theory (DFT), time-dependent DFT (TDDFT), phonon dispersion and DFT-coupled quantum transport studies, we have highlighted the interesting correlated electronic prospects of Au-Ag FCC bimetallic heterostructures. While we restrict the main-text investigations on Au-Ag systems, in the supporting information [25], we have predicted another promising correlated-electron FCC combination, *viz*. Pt-Pd. We concentrate on the [111] cleaved surfaces, sharing hexagonal topology for the corresponding Brillouin zone (BZ) for all constructed systems, as is also seen by experimental findings[1]. Procedural details of the calculations are enlisted in the supporting information [25-34]. The next section represents the DFT-derived electronic structure of different constructed geometries followed by the calculated Fermi-surfaces in the consecutive section. Next, the dynamical optical responses of the systems are calculated by TDDFT. The phonon dispersion and the quantum transport properties of the systems are presented in the successive sections.

# 1. Electronic band structure in hexagonal geometry: DFT

## 1.1. Structural construction

We have constructed the [111] cleaved *trilayer* of Au and Ag with the lattice parameter $\frac{a}{\sqrt{2}}$ in a hexagonal BZ, where *a* is the lattice parameter of their respective FCC unit cell. The electronic band structures are analyzed for five systems including the parent trilayers, *viz*, 1) Ag-111 matrix, 2) Au-111 matrix, 3) large-area interface formed from (1) and (2) having lattice mismatch of 0.12% (Au/Ag-111), 4) Ag-cluster of ~1nm embedded in Au-matrix and 5) Ag-doped Au matrix. System (4), where a few-atoms cluster (~ 1 nm in average diameter) of Ag is embedded in an Au-111 trilayer surface matrix, is a 3D NS system, whereas for system (5), the 2D system of Au-matrix is doped centrally with a few Ag atoms. To circumvent the *z*-directional periodic interaction between replica images, a vacuum layer of 20Å thickness was added between them. All these systems are checked for their surface energy convergence and ground state stability. The last row of Fig 1 depicts the structures of the systems 1-5.

## 1.2 Chemistry of correlated behavior: presence of localized bands

Fig 1(*d*) – (*k*) delineates the orbital-projected spin-orbitally coupled (SOC) electronic fatbands for all the five structurally relaxed systems along high-symmetry directions within the hexagonal BZ, as depicted in Fig 2(d). The first column of Fig 1 represents the converged charge densities for systems 3, 4 and 5. The outcomes of the investigation of electronic band structure are summarized as follows:

A) In contrast to the bulk bands, [111] surfaces of both Ag and Au have conceded presence of flat-bands slightly above the Fermi-energy ($E_F$) between X-Y, manifesting mostly *p-d* (Fig 1(d)) and *s-d* (Fig 1(e)) hybridized characters respectively. The filled 4*d* levels of Ag [111] are ~ 0.7 eV deeper than the Au-5*d* levels.

B) In Au/Ag-111 (Fig 1(f) and (g)), electrons from near- $E_F$ Au *s-d* hybridized levels are transferred to the *p-d* hybridized levels of Ag, leading to an upshift for the Au-5*d* levels between the energy range 0 to -2 eV, earlier empty for systems 1 or 2. The nature of charge transfer between Au and Ag will be evident from the Bader charge table presented at Supporting Information [25], mainly pin-pointing towards *p*-type doping for Au. Near – $E_F$ bands are mostly *p-d* hybridized for both Ag and Au and the system, being devoid of localized flat-bands due to inter-layer electron transfer from Au[111] to Ag[111], retaining the individual Fermi-liquid (FL) nature of Au or Ag. Absence of flatbands and the resulting FL nature indicate the disappearance of correlated electron behavior, as is also seen experimentally for large-area interfaces[35].

C) For the embedded NS, as is evident by comparing the energy scales of first two and last two band columns of Fig 1, there is a decrease of band width for near-$E_F$ bands. The flatter bands near $E_F$ from X to Y consist of all three orbital characters. The flatness induced correlated behavior suggest a possibility of driving the system into the non-Fermi liquid (NFL) regime. These bands also instigate the system to lose its inherent mirror symmetry of the band-structure, as existent in systems 1-3, with a Γ-centered mirror plane passing though the hexagonal vertex. Fig S4 [25] displays the PDOS of systems 3 and 4, indicating contribution of the partially-filled Au-5*d* levels at $E_F$.

D) The symmetric band-dispersion, as in system 1-3, is recovered in system (5) (Fig 1(j) and 1(k)), where the 4*d* and 5*d* derived carriers are highly localized from Γ-X and Γ-Y, suggesting a NFL nature. Interestingly, the band-topology within X-Y, having *s-p* hybridized levels, exhibits almost linear dispersion with an approximate electron-hole symmetry depicting Dirac-cone like feature near $E_F$. The band-structure clearly indicates the possibility of minute fluctuation or perturbation to be capable of opening up a gap.

E) For both systems (4) and (5), highly localized electrons in flat-bands may be responsible for promoting NFL nature, as also observed in hole-doped cuprates[36,37], where localized flat-band induced NFL behavior is realized for a significantly large region of the BZ. Table 1 presents the average effective masses of carriers for all the systems in their respective flat-band regions, a comparison of which reveals an order of magnitude increase of the effective masses for systems 4 and 5.

Thus, nanostructural morphology induced carrier-localization and orbital chemistry is capable of promoting the correlated NFL nature for these systems.

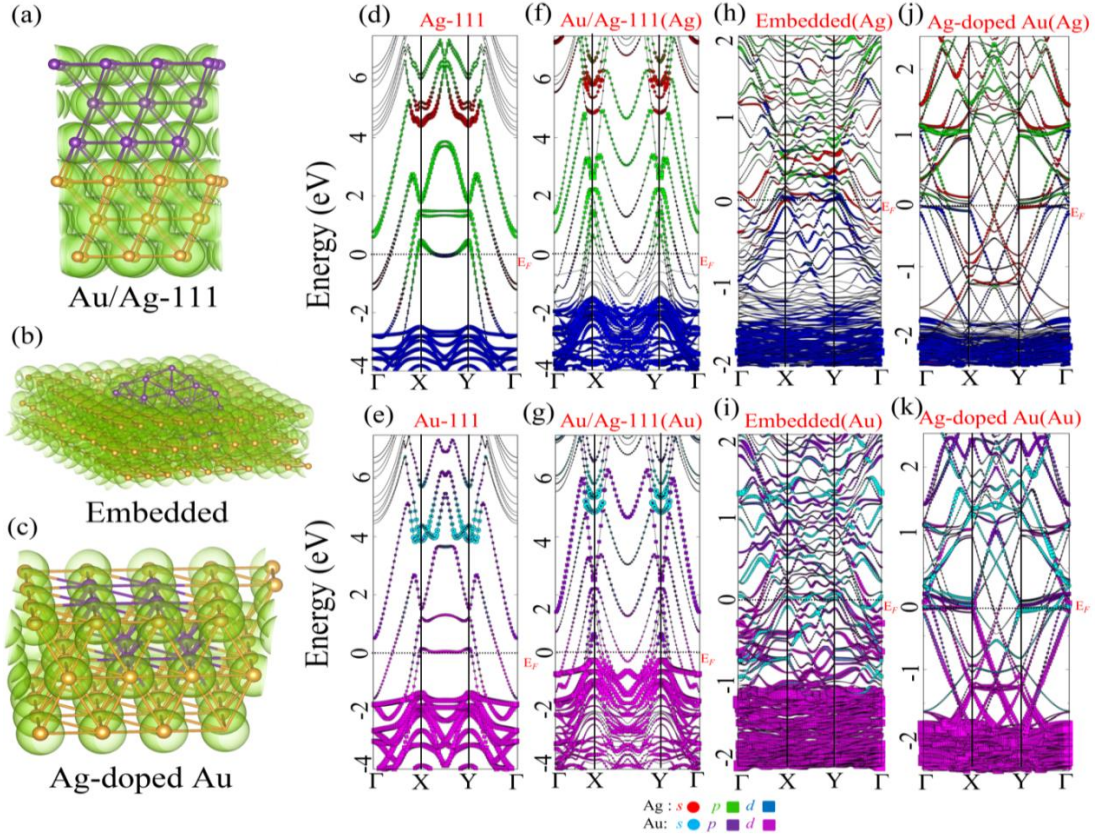

*Figure 1:* (color online) Converged charge densities for a) Au/Ag-111, b) Embedded, c) Ag-doped Au systems. The orbital projected fatbands for d) Ag-111 e) Au-111, orbital projection of f) Ag in Au/Ag-111 system, g) Au in Au/Ag-111 system, h) Ag in Embedded system and i) Au in Embedded system, j) Ag in Ag-doped Au system and k) Au in Ag-doped Au system with the respective orbital characters denoted in figure. The lowest highlighted row depicts the structures 1-5.

## 2. Fermi-surface nesting and possibility of CDW

The interdependence of Fermi-surface nesting (FSN) and the presence of charge/spin-density waves (CDW/SDW) in metals have originated from Peierl's formulation of instability for 1D periodic half-filled system[38]. Presence of instabilities like CDW/SDW conceives new periodicity within an electronic system either compatible (commensurate) or incompatible (incommensurate) with the lattice periodicity [39]. In presence of any perturbation from ground state[40], the occupied (***k***) and unoccupied (***k′***) electronic levels interact to generate a superposed density wave of periodicity defined by ***q.R***, where ***q*** = ***k-k′*** and ***R*** represents direct lattice position. While doubly occupied density wave orbitals represents CDW, SDW constitutes orbitals, singly-occupied with different spin states. In presence of a FSN, the levels corresponding to electron (occupied) and hole (unoccupied) pockets connected by ***q*** are almost degenerate for all *k* near a FS sheet. For metallic systems with reduced dimensionality, a nested FS poses an indication to the presence of CDW/SDW, whereby the nested portions of FS are removed by opening a gap. A recent reclassification[41-43] of CDW implies that for type I CDW, purely electronic FSN can be the origin of CDW and lattice distortion can be generated as an after-effect. For type II, CDWs are driven by electron-phonon coupling (EPC). To obtain the FSN-driven CDW, reduced dimensionality is an essential criterion to generate a perfectly diverging response function. For such systems, whenever symmetry is broken macroscopically by doping or other means, the FS undergoes a reconstruction forming electron and hole pockets, embodying closed orbits for quasiparticles [44]. In addition to CDW/SDW, there are contradicting studies about the FSN to be sole responsible for generating spin/orbital fluctuation-induced unconventional superconducting pairing in Fe-pnictides[45-47].

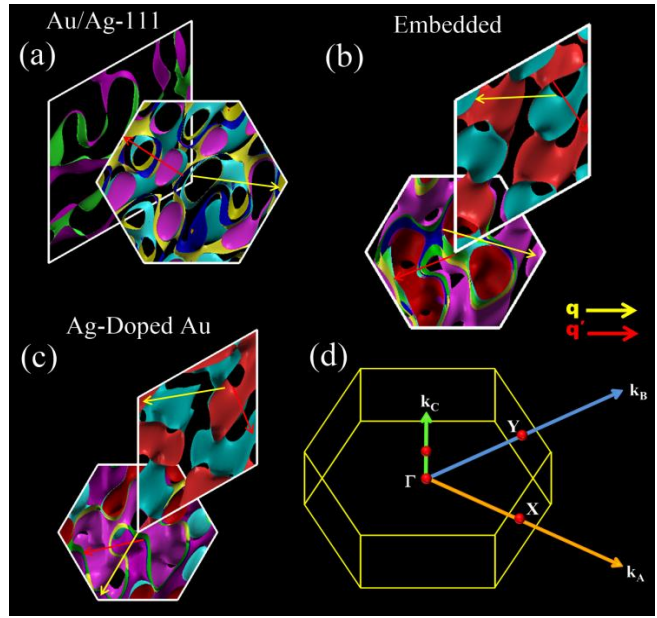

***Figure 2:*** *(color online) Merged Fermi Surface sheets corresponding to all bands crossing $E_F$ for a) Au/Ag-111 b) Embedded, c) Ag-doped Au systems within the hexagonal BZ. The rhombus represents the unit cell in the reciprocal lattice, d) Hexagonal BZ with high-symmetry directions.*

Reliance of perfection of nesting on the superconducting pairing strength[48] have motivated the analysis of FS topology for systems (3), (4) and (5). The resultant FS, for all these three cases, is complicated and composed of several individual sheets containing the electron and hole pockets representing huge FS reconstruction. Fig 2 depicts the 2D projection of the merged FS sheets within the BZ for all three cases, whereas the bottom/top superposed rhombus represents a single

FS sheet within the *k*-space unit cell. The tip of the rhombus within the BZ designates the Γ-point. The key features of the analysis of FS can be listed as follows:

A) In Fig 2(a), for Au/Ag-111, the Γ-centred hole pocket nests partially to the small electron pockets at the corner of the BZ with $q$ (yellow arrow). There is another incomplete nesting designated by $q'$ towards some mid-sides of the BZ (red arrow). Partial nesting and electron-hole asymmetry leads to the inequality of size/shape of electron and hole pockets and also connote the systems lacking potential for CDW.

B) For the embedded system (Fig 2(b)), FSN is more complicated due to the *z*-directional dispersion of the sheets. Near Γ, the hole pockets nest with electron pockets at the mid-side ($q'$) and also at the corners ($q$).

C) In Fig 2(c), due to the 2D nature of the doped system, the Γ-centred hole pocket nests perfectly with all the electron pockets situated at the corners ($q$) and also at the mid-sides ($q'$).

The reconstruction of FS and nesting-like features suggests that system 4 and 5 may be promising solicitants for CDW/unconventional superconductivity. However, in all these systems, the FSN is incommensurate in nature, with the nesting vectors $q = \frac{a}{\sqrt{2}}$ and $q' = \frac{\sqrt{3}a}{2\sqrt{2}}$, with *a* being the lattice parameter of the original lattice. The periodic replication of the FS, retaining the hexagonal symmetry, is also tested in the extended BZ in the 2D plane.

## 3. Time-dependent DFT: near zero behavior of dielectric response function

For metallic systems, in addition to the inter-band transitions, a finite probability of intra-band transitions leads to the so-called Drude-like term at the long-wavelength limit. In this section, the dynamical dielectric response and the electron-energy loss (EELS) function is computed for systems 1-5 after incorporating both intra- and inter-band orbital transitions by using TDDFT modeling to obtain an idea about the low-energy optical and plasmonic properties of these systems and their interconnection with the orbital chemistry. The details of TDDFT parameters are described in reference [25].

## 3.1 Calculation of Dynamical Response

Calculation of dynamical optical properties involves many body purturbative approach to the solution of Bethe-Saltpeter equation (BSE) using the single particle Green's function, where the dielectric function is written in terms of the exchange-correlation (xc) kernel as: $\epsilon_{GG'}^{-1}(q,\omega) = \delta_{GG'} + v_{GG'}(q)\chi'(q,\omega)$, where $v_{GG'}(q)$ is the bare Coulomb potential, a function of wave vector $q$ and $\chi'(q,\omega)$ being the full response function dependent of $q$ and frequency $\omega$. $\chi'(q,\omega)$ is related to the non-interacting density-density response function $\chi^0(q,\omega)$ as $\chi'(q,\omega) = \frac{v_{GG'}(q)\chi_{GG'}^0(q,\omega)}{1-[v_{GG'}(q)+f_{xc}^{dyn}(q,\omega)]\chi_{GG'}^0(q,\omega)}$, where the frequency-dependent dynamical exchange-correlation kernel is $f_{xc}^{dyn}(q,\omega) = -\frac{1}{q^2}(\alpha + \beta\omega^2)$, with $\alpha$ and $\beta$ being system-dependent parameters. Details of the inter- and intra-band calculations are described in the Supporting Information [25,49-53]. The calculated real ($\varepsilon_1$) and imaginary ($\varepsilon_2$) parts of $\varepsilon$ are presented in Fig 3, along with the corresponding refractive index and the dynamical electron energy loss spectra (EELS), as calculated from $\varepsilon_{loss} = -\Im[\epsilon_{GG'}^{-1}(q,\omega)]$.

## 3.1 Analysis of dynamical Response and EELS

The present subsection describes the dynamical responses for all the five systems, its interdependence with chemistry of localized band-carriers followed by the corresponding implications towards the intriguing optical responses with the help of a comparative analysis. The key findings of the time-dependent response can be listed as follows:

A) We have calculated the plasma frequencies for all these systems from TDDFT calculations, which are presented in table 1. Plasma frequency is having an inverse relationship with effective mass, as described by equation (12) of Supporting Information [25]. In comparison to large area surfaces and interfaces, as in system 1-3, an order of magnitude decrease of plasma frequency for systems 4 and 5 indicates the increase of effective mass and thereby the localization-induced NFL behavior, which is at par with the DFT prediction of localized orbital chemistry.

B) The real and imaginary parts of dynamical dielectric response functions, as computed by TDDFT, are plotted in first and second columns of Fig 3 respectively. For systems 1-3, the effects of intra-band transitions promotes the Drude-like near zero negative and positive divergence of $\varepsilon_1$ and $\varepsilon_2$ respectively. Away from zero energy, the inter-band transition peaks prevail. A close comparison of $\varepsilon_1$ and $\varepsilon_2$ for all these systems clearly indicates the fascinating optical potential of the systems 4 and 5 compared to the other systems. For both of these systems, $\varepsilon_1$ and $\varepsilon_2$ undergoes a transition to a minuscule value in the UV-vis energy range, after a near-zero positive divergence. This happens because of the massive suppression of the intra-band transitions due to localization of carriers for systems 4 and 5, as is also seen from the DFT results. Thus, for these two NS systems, flat-band induced NFL nature, presence of localized carriers with higher effective masses, reduction of plasma-frequency by an order of magnitude and epsilon near-zero (ENZ) behavior in the UV-vis range go hand in hand to demonstrate the importance of interface morphology on the orbital chemistry and the subsequent electronic properties. Two simple Drude-like systems like Au and Ag, when combined into a specific NS, suppresses all intra-band effects due to localization of carriers.

C) Such behavior endorses propitious usage of these NS in the field of non-linear optics, where a gigantic search for ENZ materials are underway[54,55]. These systems may have a large non-linear refractive index and thus may be used for optical switching. The near-perpendicular exit of transmitted light also prompts their use as optical interconnectors. For the Pt/Pd system, as presented in Supporting Information [25], ENZ values are plausible for both large-area interfaces and doped systems.

D) The fourth column of Fig 3, representing the time-dependent EELS, reveals the dynamics of both low-energy intra-band and high-energy inter-band plasmons. The surface plasmon inter-band resonance peaks for Au and Ag are well-known to get shifted towards higher energies for thinner samples [56,57]. For the trilayer Au and Ag [111] surfaces, the inter-band plasmon peaks are at ~ 3 and 2.8 eV respectively. The insets show the intra-band plasmonic transitions, which, as expected, are having peaks around the calculated plasma frequencies in Table 1. Interestingly, for systems 4 and 5, all intra-band plasmonic oscillations are frozen and a single peak of inter-band plasmon has emerged. Such frozen intra-band plasmons for artificially engineered heterostructure tally well with the highly localized electronic band-structure and diverging properties of dynamical response, where intra-band effects are seen to get suppressed for systems 4 and 5. It may be mentioned in passing that the experimental study has also observed a suppression of plasmonic resonances for the nanostructured assembly [1].

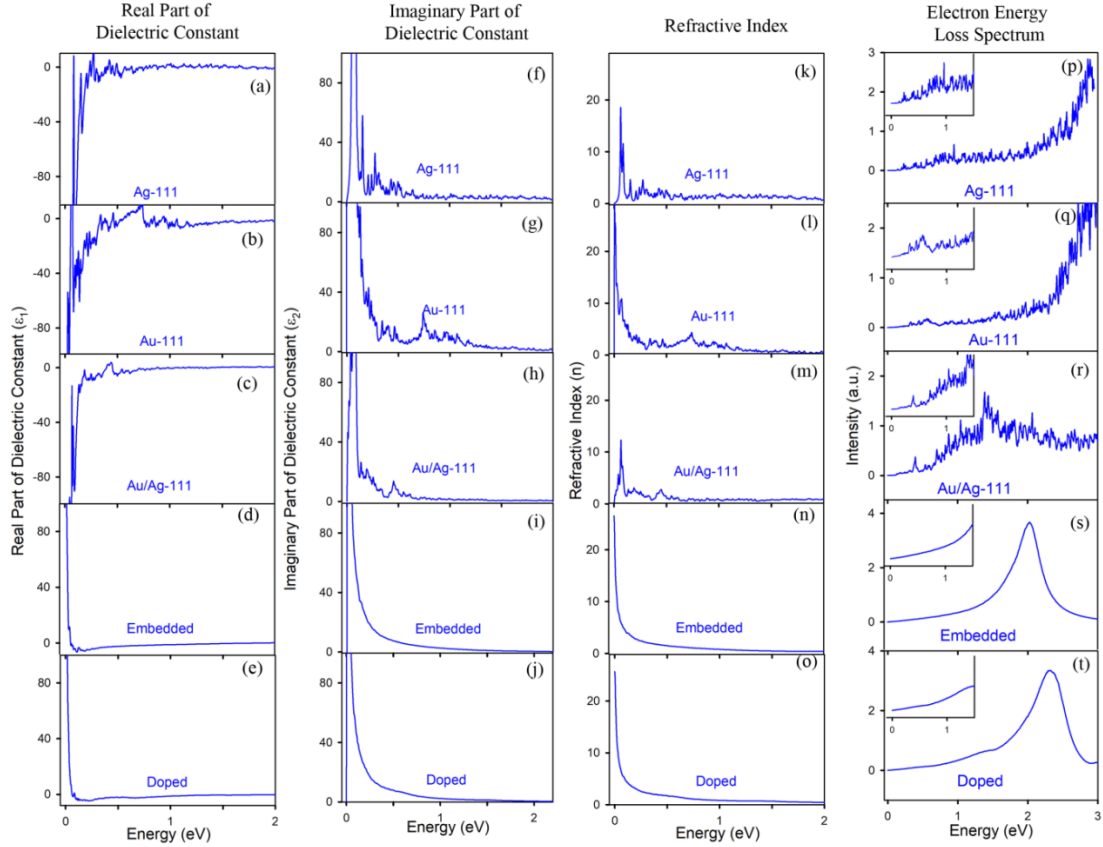

*Figure 3: (color online) Real ($\varepsilon_1$) and imaginary ($\varepsilon_2$) parts of dielectric constant ($\varepsilon$), refractive index (n) and EELS plot for different systems.*

## 4. Phonon instabilities under Ag substitutional doping

In order to have an insight about the occurrence of instabilities within the Au/Ag combined system, we have investigated the phonon dispersions and the corresponding density of states for four model systems, *viz.*, (I) infinite Au [111], (II) infinite Ag [111], (III) infinite Au [111] with ~ 33% substitutional doping of Ag (Low) and (IV) same system with ~ 66% Ag-doping (High). By infinite surface, we imply the absence of vacuum layer on top of the [111] cleaved surface. For all of these four systems, after optimization of the ionic positions and the lattice parameters, the phonon dispersion and DOS along with the corresponding electronic band dispersion are conferred in Fig. 4, the key observations of which can be indexed as below:

A) Unlike the first two systems, the doped systems demonstrate presence of soft phonons, as will be evident from the phonon band dispersion and the corresponding DOSs in Figures 4(c), 4(d), 4(g) and 4(h), implying the presence of instability in these systems[58,59]. Presence of dynamical instabilities is observed along the full high-symmetry directions Γ-X-Y-Γ, having the maximal negative energy modes at Γ point, indicating that the instability may have originated at the Γ point. The imaginary phonon modes with a negative frequency may be indicative of a structural metastability present in the system, where permanent displacements of the ionic positions may be achieved through the lattice distortions[60].

B) Although, the nested FS, as presented in section II, indicates an unconventional pairing, the calculated electron phonon coupling constant (EPC) has shown an order of magnitude change, as also experimentally observed for the doped cuprate systems[61]. The calculated EPC for system I, II and III are respectively 0.03, 0.002 and 0.68. The technical details of the phonon calculations are listed in the Supporting Information [25].

C) The phonon DOS for both of the doped systems manifests localized peaks $\sim \pm 10$ meV, indicating the participation of the corresponding modes in coupling with the charge carriers.

D) The electronic band-dispersion of the same system (Figures (i) – (l)), suggest an idea nature of charge carriers near $E_F$. For systems (I) and (II), the near-$E_F$ flatness between X and Y are due to the *p-d* hybridized levels of Ag and Au respectively, as is also seen for their finite counterparts in Fig 1. For both systems (III) and (IV), incorporation of Ag in the system leads to a switching of the nature of the near-$E_F$ carrier, associated mostly to highly localized Au *s* and *d* levels. The near-$E_F$ Ag-levels are filled up due to the charge transfer from Au. For system (IV), all three bands crossing the $E_F$ are doubly degenerate with simultaneous opening of a gap at $\sim$ 1eV. This gap, although invisibly small, is also present in system (III). While comparing the soft phonons in both of these doped cases, the phonon bands are highly localized in the second case between X to Y with more softness with the increasing extent of Ag doping.

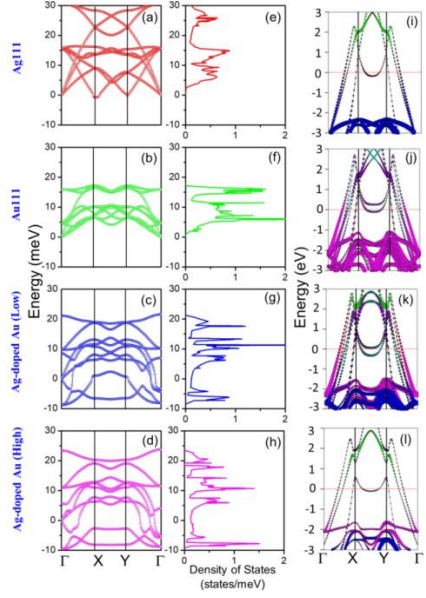

*Figure 4:* (color online) Phonon dispersion for a) Au111, b) Ag111, c) Ag-doped Au (low) and d) Ag-doped Au (high) systems. Corresponding phonon DOSs are shown in e), f) g), h) respectively. Electronic fat-bands of i)Au111, j) Ag111, k) Ag-doped Au (low) and l) Ag-doped Au (high) system with the similar legends for orbital fatness as Fig 1.

## 5. Quantum transport of Au/Ag devices

The resistivity measurements performed in reference[1-3], have persuaded the reconnaissance of the room-temperature DFT-coupled quantum transport behavior of the simplest possible device geometries made out of these two metals, where the trilayer surface of Au(Ag) will constitute a channel with lateral contacts of Ag(Au) (see Supporting Information[25]). We designate them as I) Au channel with Ag contact and II) Ag channel with Au contact. The key findings of the quantum transport calculations on both of devices can be enlisted as below:

A) In Fig 5(a), the I-V transport characteristics are plotted, implying that the absolute current value will be higher for the device II with a maximum bias of upto 2V across the contacts. The transport does not have any spin-dependence.

B) More details of transport behavior could be seen from Fig 5(b), where we plot the 2D projection of the local DOS (LDOS) in the *b-c* plane passing through the center of the device-BZ (Γ point) with respect to the current-transport (c) axis. Comparison of the details of LDOS suggests that device I, displaying more DOS at channel, have lesser interfacial backscattering at the contacts than device II resulting lesser DOS at contacts. Contribution from the incident and reflected states from the contact-channel interfaces will also be evident from the 3D LDOS plots (Fig 5(c)), which supports the 2D LDOS features of having more interfacial scattering for the device II.

C) The nature of electrical transports will be more elusive from the interpolated transmission contour plots through the central *a-b* plane (perpendicular to the current transport), passing through the Γ-point for different biases along with the total integrated transmission spectra with respect to energy. For Au-channel (device I), with increasing bias, highly transmitting zone around Γ-point starts depleting, while the edges transmits more (Fig 5(e), (f) and (g)). The integrated transmittance plot with respect to energy (Fig 5(d)) also demonstrates the reduction of total transmission through the central zone with increase in the applied bias. For device II with Ag channel, on the other hand, the transmission color-maps contain more intricacies (Fig 5(i), (j) and (k)). In addition to the depleting central zone, highly transmitting pockets are created over the Γ-centered cross-sectional area of the device.

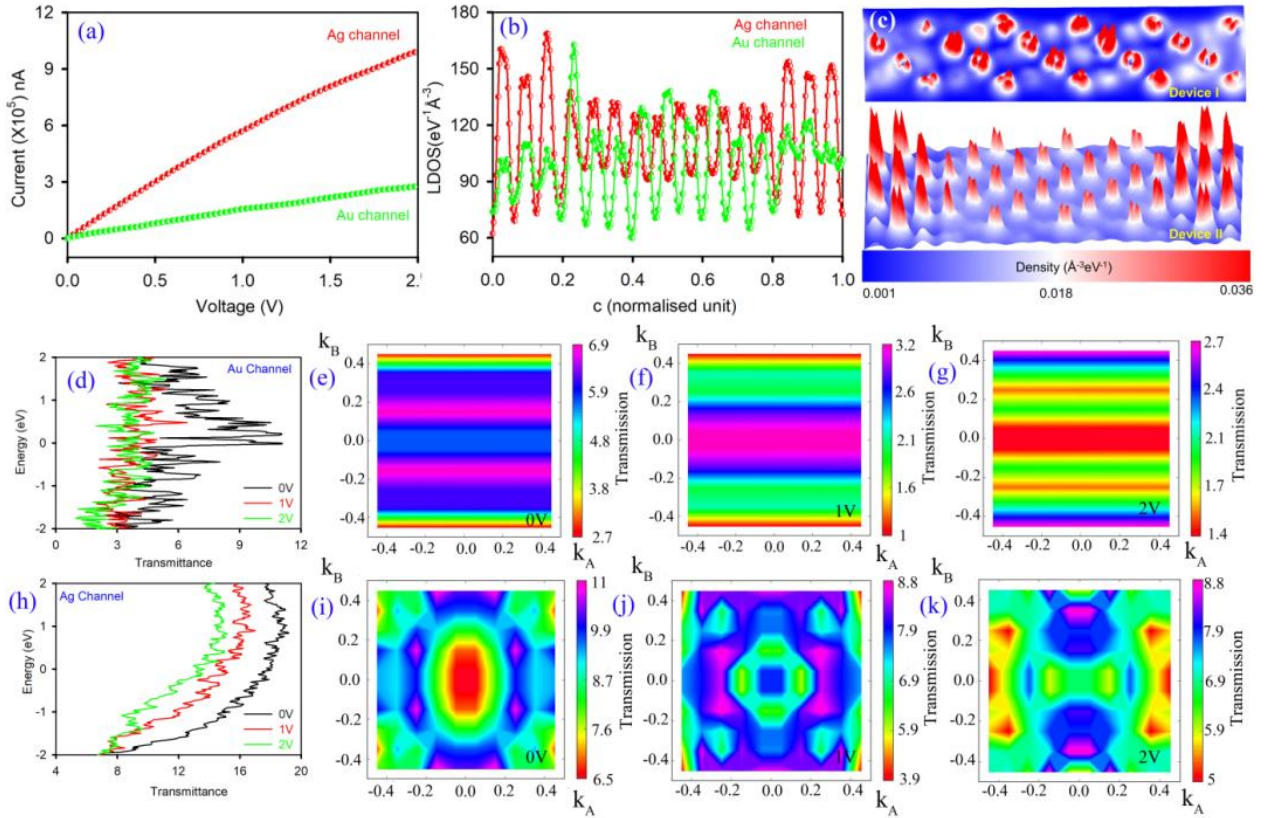

*Figure 5:* (color online) a) Comparative I-V characteristics for I) Au channel with Ag contact and II) Ag channel with Au contact, b) 1-D projected LDOS for device I and II and c) 3D coloured plot for LDOS of device I and device II. Transmittance plot for d) device I (see text) and corresponding interpolated Γ-

*centered contour plots corresponding to bias voltages e) 0V, f)1V and g) 2V, transmittance plot for h) device II and corresponding interpolated contour plot with respect to bias voltages i) 0V, j)1V and k) 2V.*

D) With increasing bias, the regions with different transmissions are reorganized, keeping the trend of having higher transmission at edges. For highest bias, even the edges contain some low-transmission region. The integrated transmittance plot in Fig 5(h) shows the clear reduction of transmission through the central plane with bias. Therefore, the transport details for the simplest possible device are far more complicated than anticipated. Experimental NS systems [1-3] will contain much more complexities because of the presence of defects, grain boundaries and different sample morphologies. The interfacial and defect scattering effects will complicate the scenario by many orders.

## 6. Conclusion

We have analyzed the electronic, optical, vibrational and transport attributes and their interdependence with the correlated orbital chemistry of nanostructured Au-Ag systems to obtain a clear understanding of the current experimental scenario. This elaborate study has the following significant results: A) DFT + SOC calculations reveal the interconnectivity of localized orbital chemistry and flat-band induced NFL nature for the embedded or doped NS, which may promote the strongly correlated phenomena in such NS, as was also seen in experiments; B) presence of complete FSN for such NS indicates the possibility of presence of CDW/SDW type instabilities; C) TDDFT- derived dynamical optical responses indicate near-zero values of the dynamical dielectric response in the UV-vis energy range for those specific NS; D) The calculated EELS for the nanostructures stipulate the presence of frozen low-energy intra-band plasmonics; E) Phonon-dispersion curve of model systems indicates the presence of dynamical instabilities in the doped system; F) DFT coupled-quantum transport results suggest that even the simplest possible device made out of these two metals have many complexities in their transport properties. Thus, this elaborate study emphasizes the interdependence of nano-structural morphology, localized orbital chemistry, higher carrier effective mass, prediction of FSN-induced CDW/SDW, near-zero epsilon behavior, suppression of intra-band plasmonics and complex quantum transport to highlight the fact that for selective NS morphologies, localized orbital chemistry may lead to the exotic electronic properties widely different from the individual components. Comprehensive investigations on more such systems may generate an altogether interdisciplinary interest, connecting physical chemistry to physics and material science and more.

**Table 1:** Effective masses of carriers (DFT) and plasma frequencies (TDDFT) for different systems are presented. The inverse relationship between these two quantities is evident.

| System | Effective Mass (DFT) | Plasma Frequency (TDDFT) (eV) |
|---|---|---|
| Ag111 | 0.4 | 0.7 |
| Au111 | 0.5 | 0.5 |
| AuAg111 | 0.06 | 0.3 |
| Embedded | 1.17 | 0.04 |
| Doped | 2.89 | 0.06 |

**Acknowledgements:**
T.K.M. acknowledges the support of DST India for the INSPIRE Research Fellowship and SNBNCBS for funding. D.K. would like to acknowledge the BARC ANUPAM supercomputing facility for computational resources, BRNS CRP on Graphene Analogues for support and motivation. We also acknowledge discussions with Shouvik Dutta from IISER Pune, Kausik Majumdar from IISC, Bangalore and Kantimay Dasgupta from IIT, Bombay, Manish Jain and Arindam Ghosh from IISC Bangalore.

**Supporting Information**
The supporting information contains the methodology of the computational calculations and theoretical overview of optical calculations. It also contains converged spin density, orbital projected density of states of different Au-Ag systems. The orbital projected fatbands and the real and imaginary part of dynamical response function of Pt-Pd system are also listed in the supporting information article.